\documentclass[12pt]{iopart}

\usepackage{graphicx}
\usepackage{epstopdf}
\usepackage{iopams}  
\usepackage[latin1]{inputenc}
\usepackage[T1]{fontenc}
\usepackage[english]{babel} 
\usepackage{array}
\usepackage{lmodern}
\usepackage{siunitx}
\usepackage[parfill]{parskip}
\usepackage{multirow}
\usepackage{eqnarray}

\begin{document}

\title{Two-mode and dual-resonant planar photonic waveguides for efficient guiding and trapping of atoms}

\author{Yuri B. Ovchinnikov}

\address{National Physical Laboratory, Hampton Road, Teddington TW11 0LW, UK}
\ead{yuri.ovchinnikov@npl.co.uk}
\vspace{10pt}
\begin{indented}
\item[]31 July 2025
\end{indented}

\begin{abstract}
The trapping of ultracold atoms using two-colour evanescent light waves formed by propagating modes of suspended optical rib waveguides is modelled in different configurations. Reducing the anisotropy of the two-colour evanescent optical dipole potential requires two laser light components with a large frequency difference. The upper frequency is guided in the two lowest transverse waveguide modes and the lower frequency propagates in a single-mode regime. This increases the dipole potential depth in the lateral direction. An additional increase in the optical dipole potential can be achieved by tuning the frequencies of the two modes to two different atomic transitions. When applied to rubidium, the total depth of the corresponding surface optical dipole traps can reach 0.3 mK or above under reasonable conditions.

\end{abstract}

%\maketitle

\section{Introduction}
All-optical atom chips \cite{1} for coherent manipulation of ultra-cold neutral atoms based on atom-photonic waveguides \cite{2} have great potential for quantum sensing \cite{3}, on-chip atomic clocks, and quantum computing with optically trapped neutral atoms \cite{4}. Initially, the trapping of atoms in two-colour evanescent light waves was proposed using two laser beams undergoing total internal reflection at the interface between dielectric media and vacuum \cite{5}. This configuration allows the transverse sizes of the laser beams and their angles of total reflection to be chosen freely, making it easy to realise stable trapping of atoms in two-colour evanescent light fields, as demonstrated in \cite{6}. Another configuration for trapping atoms in two-colour evanescent light fields uses nanofibres \cite{7,8}.
The cylindrical shape of a nanofibre eliminates the lateral stability problems of the corresponding optical dipole traps, which can have a depth of up to 0.4 mK \cite{9}.

The use of planar photonic waveguides for trapping atoms in two-colour evanescent light waves presents two major challenges. For a standard planar photonic waveguide fabricated on top of a dielectric substrate with a smaller refractive index, the maximum propagation angle for supported modes is limited by the difference in refractive index between the waveguide and the substrate. This reduces the penetration depth of the evanescent light waves in a vacuum, where the atoms are supposed to be trapped. In the presence of van der Waals interactions between the atoms and the dielectric waveguide, this results in the requirement of unrealistically high laser powers in the two-colour modes of the waveguide in order to trap the atoms. This problem can be solved by using suspended planar optical waveguides \cite{10}, which provide evanescent light field penetration depths similar to those of nanofibres. Another problem is that, for all types of planar photonic waveguide, the higher-frequency mode has a smaller lateral size than the lower-frequency mode. This is the opposite of the condition required for stable lateral trapping of atoms in a corresponding two-colour evanescent optical dipole trap. Fortunately, under certain conditions, the lateral trapping of atoms in such potentials is still possible, albeit much weaker than the trapping of atoms in a direction normal to the waveguide surface. This large spatial anisotropy of the trap potentials typically limits their depth to several microkelvins. These reasons explain why, in spite of multiple different proposals \cite{1,2,10,11,12,13,14,15,16}, the trapping of atoms in planar photonic waveguides has never been observed to date.

This paper discusses how to significantly increase the total depth of two-colour evanescent light traps formed by the propagating modes of a two-mode suspended optical rib waveguide.

The paper is organised as follows: The second section considers the fundamental characteristics of a bi-exponential potential. The third section describes how all the trap potentials in this paper were calculated. Section four calculates the surface optical dipole potential of two modes of a single-mode suspended optical rib waveguide that are in quasi-resonance with the lowest electron transition of the $^{87}$Rb atom. Section five considers a two-mode waveguide case in which two frequency components are far-frequency detuned from the lowest electron transition of the $^{87}$Rb atom. The sixth section discusses the case of a two-mode waveguide with two frequency components that interact primarily with two different electron transitions of the $^{87}$Rb atom. 

\section{General properties of bi-exponential optical dipole potential}

A one-dimensional optical dipole potential, $U_{dip}$, formed by two evanescent light waves, which have opposite frequency detunings from a certain atomic transition can be written as
\begin{equation} \label{EQ1} 
	U_{dip}\left(x\right)=U_r\left(x\right)+U_b\left(x\right)=U_re^{\left(-2{x}/{d_r}\right)}+U_be^{\left(-2{x}/{d_b}\right)}, 
\end{equation} 
where $U_r(x)$ and $U_b(x)$ are the optical dipole potentials of the two evanescent light waves frequency detuned to the red and blue sides of the atomic transition. The $U_r$ and $U_b$ are values of the dipole potentials of the two evanescent waves at the surface of the waveguide (Fig.1), where $x=0$, and $d_r$ and $d_b$ are penetration depths of these optical evanescent fields. Note, that according to properties of optical dipole potentials \cite{1}, for the red frequency detuned evanescent wave $U_r<0$ and for the blue detuned one $U_b>0$. 

The minimum of the potential $U_{dip}(x)$ is reached at the distance

\begin{equation} \label{EQ2} 
	x_0=\frac{1}{2}\frac{d_r d_b}{d_r-d_b}{\mathrm{ln} \left( S \right) },     
\end{equation} 
where
\begin{equation} \label{EQ3} 
	S=-\frac{d_r U_b}{d_b U_r}.      
\end{equation}

For $x_0$ to be positive, $d_r > d_b$ and $S>1$. The parameter $S$ is also responsible for the depth of the trap (\ref{EQ4}). 

From (\ref{EQ2}) it follows that the ratio between the amplitudes of red-detuned and blue-detuned potentials at the minimum of the potential $x=x_0$ is determined by $-U_r\left(x_0\right)/U_b\left(x_0\right)=d_r/d_b$.

The depth of the bi-exponential potential is given by

\begin{equation} \label{EQ4} 
	U_{dip}\left(x_0\right)=U_b S^{-\frac{d_r}{d_r-d_b}}+U_r S^{-\frac{d_b}{d_r-d_b}}.
\end{equation}

The above equation can be used to derive the ratio between the depth of the bi-exponential dipole potential, $U(x)$, and the potential of the red-detuned component, $U_r(x)$, at $x=x_0$, which is determined by the penetration depths only. 

\begin{equation} \label{EQ5} 
	\eta=\frac{U\left(x_0\right)}{U_r\left(x_0\right)}=\frac{d_r-d_b}{d_r},
\end{equation}

We will refer to this parameter as the efficiency coefficient of the bi-exponential optical dipole potential and denote it by the symbol $\eta$. The subsequent sections will present examples of typical values for this parameter.

\section{Potential of optical evanescent wave trap}

In this section, we describe how to calculate the trapping potential for  $^{87}$Rb atoms in two-colour evanescent light waves near the rib surface of a suspended silica optical rib waveguide (Fig.1). The main part of the trapping potential consists of the optical dipole potentials of the two evanescent light fields, which have different penetration depths in a vacuum, different frequencies and opposite frequency detuning signs relative to certain atomic electron transitions  \cite{5}. The strength of the interaction between the induced dipole moment of an atom and the amplitude of the driving light field is characterised by the electric ac polarizability \cite{17}.

\begin{figure}[htbp]
	\centering
	\includegraphics[width=6cm]{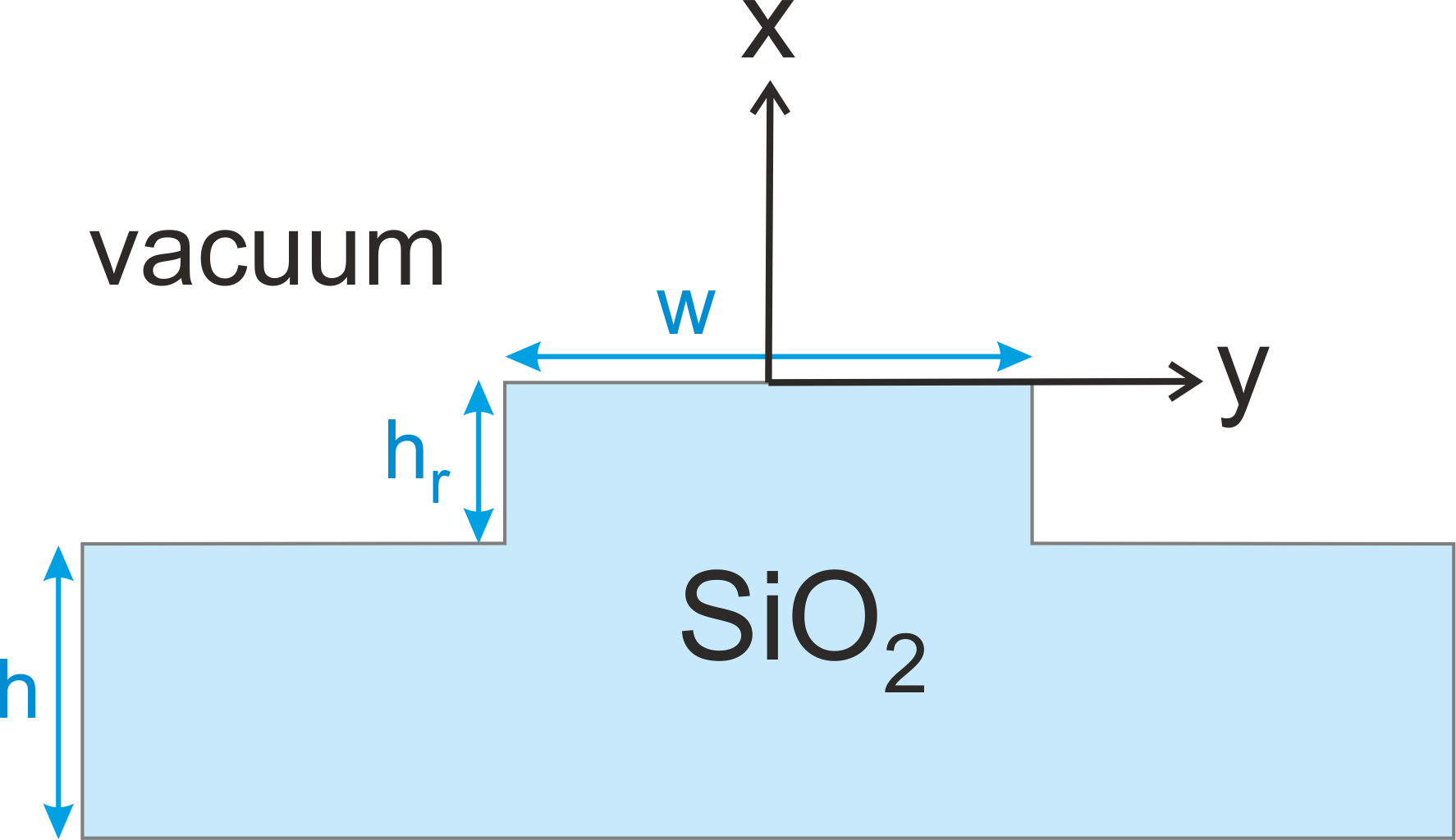}
	\caption{Geometry of a suspended optical rib waveguide.}
\end{figure}

The valence part of the scalar ac polarizability for the ground state $5^2S_{1/2}$ of $^{87}$Rb atoms driven by light of frequency $\omega$ is given by the formula \cite{18}:

\begin{equation}  \label{EQ6} 
	\alpha_v\left(\omega\right) = \frac{1}{3 \hbar}
	\sum_{n=5}^{8} \left( \frac{\omega_{np_{1/2}} \langle 5s \left\Vert D \right \Vert n p_{1/2} \rangle^2} {\omega_{np_{1/2}}^2-\omega^2}+\frac{\omega_{np_{3/2}} \langle 5s \left\Vert D \right \Vert n p_{3/2} \rangle^2}{\omega_{np_{3/2}}^2-\omega^2} \right),
\end{equation}

where $\omega_{np_{j'}}$ are frequencies of the $5^2S_{1/2}-n^2P_{j'}$ atomic transitions, $j'=1/2, 3/2$, $n=5,6,7...$, and $\langle 5s \left\Vert D \right \Vert n p_{j'} \rangle$ are reduced dipole matrix elements of the corresponding transitions. The corresponding value of the optical dipole potential is given by:

\begin{equation}  \label{EQ7} 
	U\left(\mathbf{r},\omega\right)=-\frac{1}{2\epsilon_0 c} \alpha_v(\omega) I(\mathbf{r}),
\end{equation}
where $I(\mathbf{r})$ is the light intensity at the point with coordinates $\mathbf{r}$. 

The wavelengths and values of the reduced dipole matrix elements for the $5^2S_{1/2}-n^2P_{j'}$ atomic transitions of $^{87}$Rb are presented in Table 1. The values of the dipole matrix elements for the transitions to	$5\mathrm{^2P}_{1/2}$ and $5\mathrm{^2P}_{3/2}$ were taken from \cite{19}, and the values for $6\mathrm{^2P}_{1/2}$ and $6\mathrm{^2P}_{3/2}$ were taken from experimental work \cite{20}. All other data were taken from the original theoretical modelling \cite{17}.

\begin{table}[htbp]
	\caption{Wavelengths and values of reduced dipole matrix elements of lowest allowed transitions of $^{87}$Rb}
	\label{tab:shape-functions}
	\centering
	\begin{tabular}{ccc}
		\hline
		Excited state & $\lambda$ (nm) & $\langle 5s \left\Vert D \right \Vert n p_{j'} \rangle$ (C$\cdot$m)\\
		\hline
		$5\mathrm{^2P}_{1/2}$ & $794.98$ & $3.606\cdot10^{-29}$\\
		$5\mathrm{^2P}_{3/2}$ & $780.24$ & $5.090\cdot10^{-29}$\\
		$6\mathrm{^2P}_{1/2}$ & $421.67$ & $2.743\cdot10^{-30}$\\
		$6\mathrm{^2P}_{3/2}$ & $420.30$ & $4.434\cdot10^{-30}$\\
		$7\mathrm{^2P}_{1/2}$ & $359.16$ & $9.750\cdot10^{-31}$\\
		$7\mathrm{^2P}_{3/2}$ & $358.71$ & $1.713\cdot10^{-30}$\\
		$8\mathrm{^2P}_{1/2}$ & $335.08$ & $5.002\cdot10^{-31}$\\
		$8\mathrm{^2P}_{3/2}$ & $334.87$ & $9.411\cdot10^{-31}$\\
		
		\hline
	\end{tabular}
\end{table}

Another important parameter of an optical dipole trap is the spontaneous scattering rate of the trapping light's photons by the trapped atoms. This defines the trap coherence time and heating rates. The spontaneous scattering rates were calculated using the formula:

\begin{equation}  \label{EQ8} 
	\Gamma_{sc}\left(\mathbf{r},\omega_i\right) =\sum_{n=5}^{8} \left( \Gamma_{np_{1/2}} \frac{\Omega_{np_{1/2}}(\mathbf{r})^2}{4 (\omega_{np_{1/2}}-\omega_i)^2}+\Gamma_{np_{3/2}} \frac{\Omega_{np_{3/2}}(\mathbf{r})^2}{4 (\omega_{np_{3/2}}-\omega_i)^2} \right),
\end{equation} 
where $	\Gamma_{npj'}$ is the natural linewidth of the corresponding  $5^2S_{1/2}-n^2P_{j'}$ transition, and the second term corresponds to the population of the excited $n^2P_{j'}$ state, provided that $\Gamma_{np_{j'}} \ll \Omega_{np_{j'}}(\mathbf{r}) \ll |\omega_{np_{j'}}-\omega_i|$. The frequency $\omega_i$, where $i=r,b$, corresponds to the two different frequencies $\omega_r$ or $\omega_b$ of the waveguide's modes. The term $\Omega_{npj'}(\mathbf{r})$ is the Rabi frequency of the $5^2S_{1/2}-n^2P_{j'}$ transition, which is defined as

\begin{equation}  \label{E9} 
	\Omega_{npj'}(\mathbf{r}) = \sqrt{\frac{I(\mathbf{r})}{2 \epsilon_0 c}} \frac{\langle 5s \left\Vert D \right \Vert n p_{j'} \rangle }{\hbar}.
\end{equation} 
Note that, due to the strong dependence of the scattering rates on the frequency detuning of the driving fields, it is usually sufficient to consider only the closest atomic transitions in the sum (\ref{EQ8}).

The total potential, $U_{tot}$, of the evanescent light trap, including the van der Waals interaction between the trapped atoms and the dielectric surface of the waveguide, was calculated as follows \cite{1}:
\begin{equation}  \label{EQ10} 
	U_{tot}\left(x,y\right) = U_{r}\left(0,y\right)\ {\mathrm{exp} \left(-{2x}/{d_r}\right)+U_{b}\left(0,y\right){{\mathrm{exp} \left(-{2x}/{d_b}\right) 
				-\frac{C_3{{\lambda }_{eff}}/{2\pi }}{x^3\left(x+{{\lambda }_{eff}}/{2\pi }\right)}}}},
\end{equation} 
where the first two terms correspond to the optical dipole potentials of the two evanescent wave fields, while the third term represents the interaction potential between the rubidium atoms and the dielectric surface of the SiO$_2$ rib waveguide. The van der Waals coefficient is $C_3$ = 5.699 $\times 10^{-49}$ Jm$^3$ is, and the effective wavelength is $\lambda_{eff}$ = 710 nm \cite{21}. The lateral distributions of optical dipole potentials, $U_r(0,y)$ and $U_b(0,y)$, along the surface of the waveguide rib are calculated using equations (5) and (6), based on the corresponding local mode intensities.

\section{Single-mode mono-resonant atomo-photonic waveguide}

According to equations (\ref{EQ6}) and (\ref{EQ7}),  the magnitudes of the optical dipole potentials increase resonantly near atomic transition frequencies. Therefore, tuning the frequencies of the two waveguide modes close to a particular atomic transition should enhance the optical dipole potential. In the following, two modes with wavelengths $\lambda_r=782$ nm and $\lambda_b=778$ nm, which are detuned by $\sim 2$\,nm from the  $5^2S_{1/2}-5^2P_{3/2}$ transition of $^{87}$Rb, $\lambda_0=780.24$\,nm are considered.

\begin{figure}[htbp]
	\centering
	\includegraphics[width=9cm]{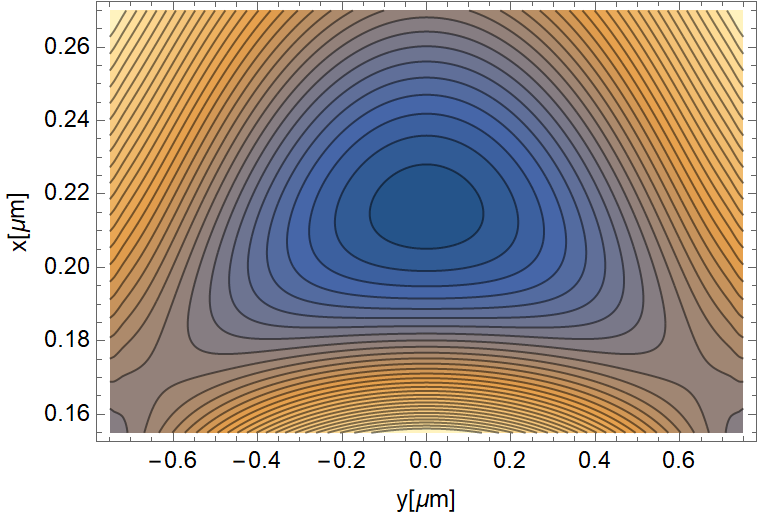}
	\caption{Cross section of the total potential for two TE$_{00}$ modes at $\lambda_r=782$ nm and $\lambda_b=778$ nm of a suspended optical rib waveguide at maximum surface light intensities $I_{r}(0,0)=6.6\times10^9$ W/m$^2$ and $I_{b}(0,0)=7.5\times10^9$ W/m$^2$.}
\end{figure}

Numerical simulations of the supported modes of a suspended silica optical rib waveguide \cite{22}, show that a waveguide with $h=300$ nm, $h_r=30$ nm and $w=1.5$ $\mu$m (Fig.\,1) only supports single TE$_{00}$ modes at these wavelengths. The penetration depths of the evanescent waves of these modes in a vacuum are $d_r=151.37$\,nm and $d_b=150.30$\,nm, respectively.  

The total potential, calculated from equations (\ref{EQ6},\ref{EQ7},\ref{EQ10}), for optical powers in the two modes of $P_r=7.3$\,mW and $P_b=8.3$ mW is shown in Fig.\,2. The maximum potential depth at its minimum, at $x_0=215$\,nm, is $U_{tot}(x_0,0)/k_B$=25.3\,$\mu$K, where $k_B$ is the Boltzmann constant. The potential depth in the lateral direction is determined by the potential depth at the saddle points: $x_s=0.17$\,$\mu$m, $y_s=\pm 0.65$\,$\mu$m. The total potential at these points is $U_{tot}(x_s,\pm y_s)/k_B=21.25$ $\mu$K. The minimal depth of the total potential is given by the lateral potential depth: $\Delta U_{tot}^{min}/k_B=(U_{tot}(x_0,0)-U_{tot}(x_s,y_s))/k_B=4$\,$\mu$K.  
The efficiency coefficient, given by equation (\ref{EQ4}), is quite small: $\eta=7.07 \times 10^{-3}$. This indicates that the individual potential depths $U_r(x_0,0)/k_B=-3.69$\,mK and  $U_b(x_0,0)/k_B=3.67$\,mK at the trap minimum are much larger than the resulting trap depth. This results in a high total spontaneous scattering rate of photons at $x=x_0$ of $\Gamma_{sc}=6200$ s$^{-1}$ (\ref{EQ8}). This scattering rate leads to the heating of the trapped atoms at a rate of approximately $dT/dt=(2/3)E_r\Gamma_{sc}/k_B=750$ $\mu$K/s, where $E_r=\hbar^2 k^2/(2m)$ is the recoil energy of a single photon at the frequency of the $5^2S_{1/2}-5^2P_{3/2}$ transition, and $k=2 \pi/\lambda_0$ is a wave vector of light at that frequency. This should limit the trapping time of atoms to less than 5 ms. Another issue is the parameter $S=1.02$. Note, that at $S=1$ the minimum of the trap is located at the surface of the waveguide. This means that even slight fluctuations or changes to the $U_b/U_r$ ratio will cause a significant change in the position and depth of the trap minimum.
Therefore, this configuration is not deep or stable enough for the efficient trapping of atoms.

\section{Two-mode mono-resonant atomo-photonic waveguide}

In this section, we consider the case of far frequency-detuned light components, when the frequency of the blue component, $\lambda_b=698$\,nm, is almost twice that of the red component, $\lambda_r=1300$\,nm. It is not possible for these light components to propagate in a single-mode regime in the same waveguide.  

\begin{figure}[htbp]
	\centering
	\includegraphics[width=8cm]{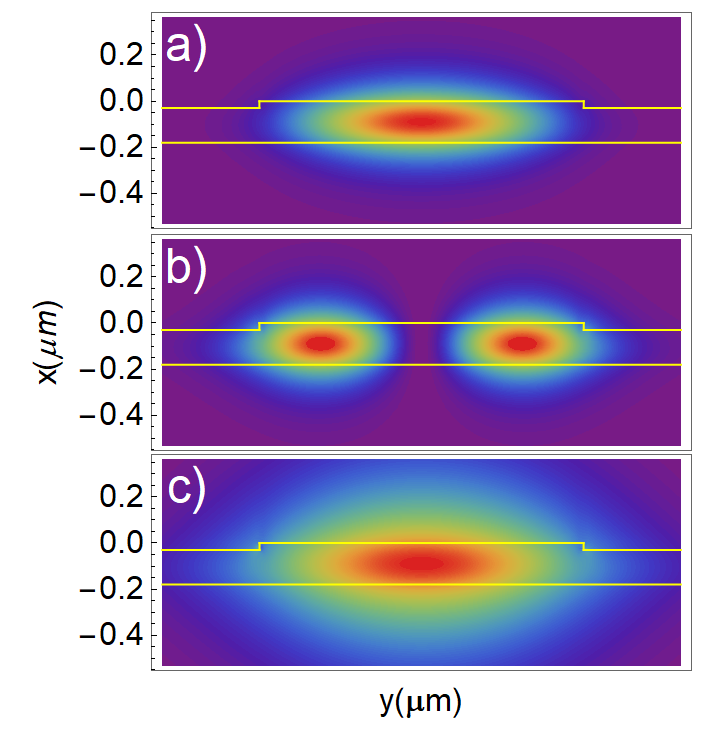}
	\caption{Numerically simulated spatial distributions of light intensities in guiding modes of the suspended optical rib waveguide in xy-plane: a) TE$_{00}$ mode at 698 nm; b) TE$_{01}$ mode at 698 nm; c) TE$_{00}$ mode at 1300 nm.}
\end{figure}

Single-mode propagation can be achieved for 1300\,nm light, whereas 698\,nm light propagates in a two-mode regime. Numerical simulations \cite{21} show that a suspended silica rib waveguide with $h=150$ nm, $h_r=30$ nm and $w=2.5$\,$\mu$m (Fig.\,1) supports the propagation of a single TE$_{00}$ mode at 1300\,nm with an effective refractive index of $n_{eff}=1.0756$, a propagation constant of $\beta=5.200$\,$\mu$m$^{-1}$, and a decay length of $d_r=488.1$\,nm, as well as two transverse modes at 698\,nm. The TE$_{00}$ mode at 698 nm is characterised by $n_{eff}=1.1917$, $\beta=10.727$\,$\mu$m$^{-1}$, $d_{b}=174.0$\,nm, and the TE$_{01}$ mode by $n_{eff}=1.1778$, $\beta=10.601$\,$\mu$m$^{-1}$ and $d_{b}=168.4$\,nm.
The simulated spatial distributions of light intensity in the three propagating modes across the suspended waveguide are shown in Fig.3.

\begin{figure}[htbp]
	\centering
	\includegraphics[width=8cm]{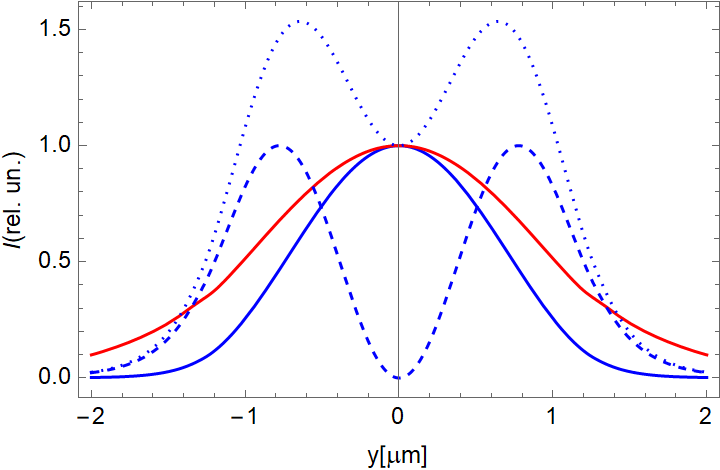}
	\caption{Normalized intensity distributions of light in the modes of the suspended optical rib waveguide along $y$-axis. Solid red curve is the TE$_{00}$ mode at 1300 nm, solid blue curve is the TE$_{00}$ mode at 698 nm, dashed blue curve is the TE$_{01}$ mode at 698 nm and dotted curve is a sum of intensities of TE$_{00}$ and TE$_{01}$ modes at 698 nm.}
\end{figure}

Cross sections of the normalised intensity distributions in the modes along the $y$-axis are are shown in Fig.\,4. The blue solid curve shows the lateral distribution of light intensity in the TE$_{00}$ mode at 698\,nm. The red solid curve shows the TE$_{00}$ mode at 1300\,nm. As can be seen, the lateral size of the blue-detuned mode is smaller than that of the red-detuned mode. This means that, for the same amplitude of the light fields, the repulsive forces generated by the blue-detuned light prevail over the attractive forces generated by the red-detuned light. This is because the lateral gradient of the 698\,nm field is larger than that of the 1300\,nm field. Fortunately, due to the different decay lengths of the two evanescent light fields, the amplitude and lateral gradient of the 1300\,nm field are larger than those of the 698\,nm field at the minimum of the trap, making the trapping stable in both transverse directions. However, this comes at the cost of a restricted central region and reduced total potential depth in the lateral direction.     

\begin{figure}[htbp]
	\centering
	\includegraphics[width=10cm]{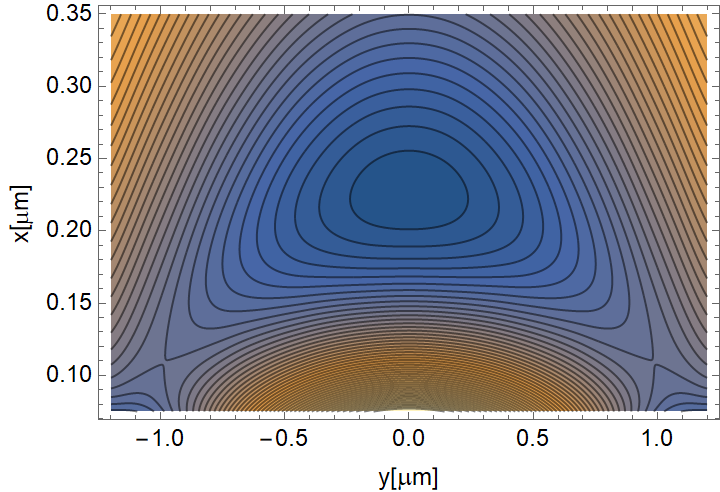}
	\caption{Cross section of the total potential for two TE$_{00}$ modes at $\lambda_r=1300$ nm and $\lambda_b=698$ nm of a suspended optical rib waveguide at maximum surface light intensities $I_{r}(0,0)=1.5\times10^{10}$ W/m$^2$ and $I_{b}(0,0)=1.2\times10^{10}$ W/m$^2$.}
\end{figure}

The total potential of the trap (\ref{EQ3}) is simulated for the two TE$_{00}$ modes, with wavelengths of 1300\,nm and 698\,nm, and corresponding total powers of 7.03\,mW and 7.73\,mW. The maximum intensities at the rib surface are $I_{r}(0,0)=1.5\times10^{10}$ W/m$^2$ and $I_{b}(0,0)=1.2\times10^{10}$ W/m$^2$, as shown in Fig\,5. The potential minimum of $U_{tot}(x_0,0)/k_B=428$\,$\mu$K is located at $x_0=228$\,nm, $y=0$, while the minimal potential depth, which is determined by the lateral depth of the potential, is $\Delta U_{tot}^{min}/k_B=99$\,$\mu$K. This is a significant depth for the trap, which can be explained by the large lateral size of the waveguide $w=2.5$\,$\mu$m, which extends the lateral sizes of the modes and the corresponding region of stable lateral trapping of atoms. The trap is characterised by the parameters $\eta=0.644$ and $S=5.38$, which determine its high efficiency and stability. The spontaneous scattering rates of atoms located near the trap minimum, estimated using formula (\ref{EQ8}), are $\Gamma_{sc}(698 \text{\,nm})=4.16$\,s$^{-1}$ and $\Gamma_{sc}(1300\text{\,nm})=2.76$\,s$^{-1}$, respectively. The heating rate of the atoms due to the spontaneous scattering of photons from the trapping light fields is approximately $0.8$\,$\mu$K/s.       

\begin{figure}[htbp]
	\centering
	\includegraphics[width=10cm]{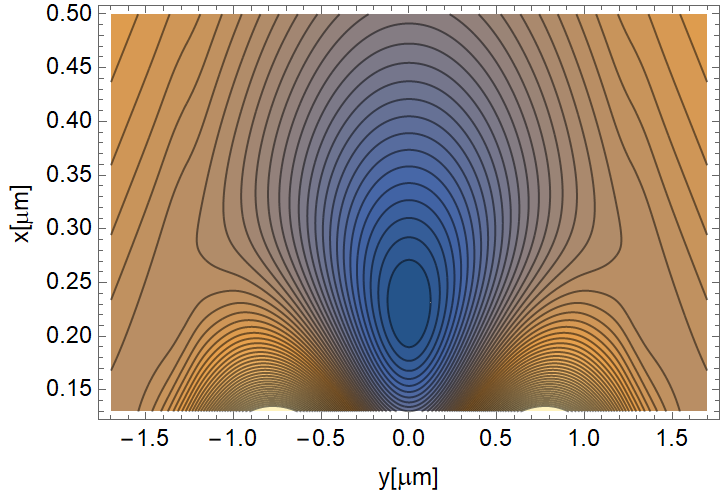}
	\caption{Cross section of the total potential for two TE$_{00}$ modes at $\lambda_r=782$ nm and $\lambda_b=778$ nm of a suspended optical rib waveguide at maximum surface light intensities $I_{r}(0,0)=6.6\times10^9$ W/m$^2$ and $I_{b}(0,0)=7.5\times10^9$ W/m$^2$.}
\end{figure}

The dashed blue curve in Fig.\,4 shows the normalised intensity distribution of the 698\,nm TE$_{01}$ mode along the $y$-axis. The maximums are located at $y_{TE01}^{max}=\pm 0.776$\,$\mu$m. The dotted blue line shows the sum of the 698\,nm TE$_{00}$ and TE$_{01}$ modes with the same amplitudes. Clearly, the lateral width of the superposition of the TE$_{00}$ and TE$_{01}$ modes is larger than that of the TE$_{00}$ mode. Therefore, adding the 698\,nm TE$_{01}$ mode is expected to improve the lateral depth of the trap. Figure\,6 shows the total potential of the trap formed by the 1300\,nm TE$_{00}$ mode with a power of 7.03\,mW, the 698 nm TE$_{00}$ mode with a power of 7.73\,mW and the 698\,nm  TE$_{01}$ mode with a power of 8.65\,mW. At these powers the maximum intensities of the 698\,nm TE$_{00}$ and TE$_{01}$ modes at the surface of the rib are equal to $1.2\times10^{10}$ W/m$^2$. The minimum lateral trap depth is increased to $\Delta U_{tot}^{min}/k_B=317.6$\,$\mu$K.    

\begin{figure}[htbp]
	\centering
	\includegraphics[width=10cm]{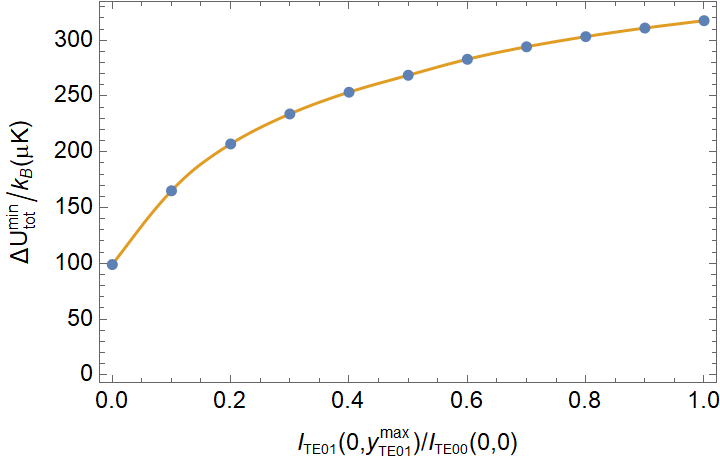}
	\caption{Minimum depth of the total potential for a trap formed by $\lambda_r=1300$ nm and $\lambda_b=698$ nm modes of a two-mode suspended optical rib waveguide as a function of relative intensity of the 698 nm TE$_{01}$ mode.}
\end{figure}

Figure 7 shows the dependence of the minimum trap depth on the relative portion of the TE$_{01}$ mode in the guided 698\,nm light. This dependence demonstrates saturation at high values of the TE$_{01}$ mode relative intensity and is more efficient at lower intensities.

It can be concluded that this trap configuration is efficient in terms of power consumption and provides a stable, deep trap for all ratios of TE$_{00}$ and TE$_{01}$ powers of 698\,nm light.   

\section{Two-mode dual-resonant atomo-photonic waveguide}

This section discusses the use of two different electron transitions in an atom to realise a two-colour evanescent light trap. The idea behind this approach is to maintain strong coupling of both frequency components of the two-colour light with the atom while the frequency difference is large. This should enable the use of a two-mode photonic waveguide to further increase the lateral trapping of atoms.

\begin{figure}[htbp]
	\centering
	\includegraphics[width=5cm]{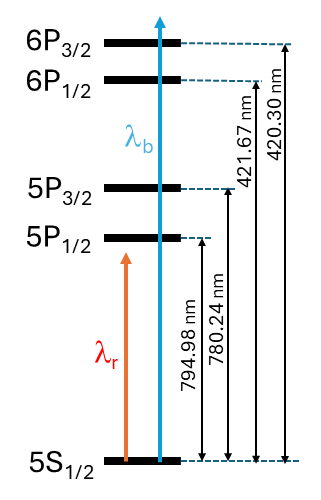}
	\caption{Energy levels of four lowest $5s-np_{j'}$ transitions of $^{87}$Rb and frequencies of the two modes of the waveguide.}
\end{figure}

The choice of frequencies for the two optical fields of the trap for $^{87}$Rb atoms is shown in Fig\,8. The frequency of the "blue" optical field component of the two-colour evanescent light trap is tuned above the $5S-6P$ transition, while the "red" component is tuned below the $5S-5P$ transition. To achieve the maximum depth of the trap, the wavelength of the "blue" component is set to equal to $\lambda_b=420.2$\,nm, which is just $0.1$\,nm detuned from the  $5^2S_{1/2}-6^2P_{3/2}$. The frequency of the "red" component is chosen to be far below the $5S-5P$ transition at $\lambda_b=980$\,nm. This choice is made to realise here the two-mode approach described in Section 5. The large difference in frequency detuning between the two optical modes is due to the different strengths of the atomic transitions involved.

According to the numerical simulations \cite{22}, a suspended optical rib waveguide made of silica  with $h=150$ nm, $h_r=15$ nm and $w=1.5$ $\mu$m (Fig.\,1) supports the propagation of a single TE$_{00}$ mode at 980 nm with effective refractive index of the waveguide $n_{eff}=1.1065$ and a propagation constant of $\beta=7.0944$ $\mu$m$^{-1}$, whereas 420.2 nm light includes two modes: TE$_{00}$ with $n_{eff}=1.2870$ and $\beta=19.2487$ $\mu$m$^{-1}$ and TE$_{01}$ mode with $n_{eff}=1.2768$ and $\beta=19.0963$ $\mu$m$^{-1}$. The corresponding penetration depths of the evanescent waves of these modes in vacuum are $d_r=313.8$ nm and $d_b=81.8$ nm, respectively.

\begin{figure}[htbp]
	\centering
	\includegraphics[width=10cm]{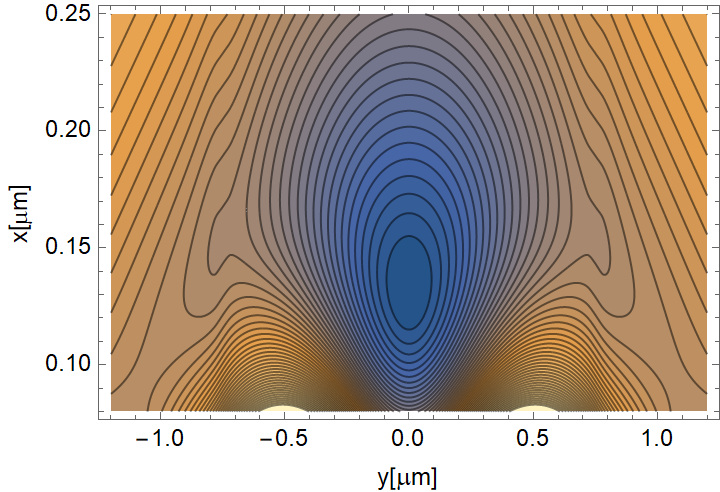}
	\caption{Cross section of total potential for one TE$_{00}$ mode of a suspended optical rib waveguide at $\lambda_r=980$ nm and two modes TE$_{00}$ and TE$_{01}$ at $\lambda_b=420.2$ nm.}
\end{figure}

The total potential of the trap for powers of modes P$_{\text{TE00}}(980 \text{nm})=6.3$\,mW, P$_{\text{TE00}}(420.2 \text{nm})=4.66$\,mW and P$_{\text{TE01}}(420.2 \text{nm})=5.78$\,mW is shown on Fig\,9. The maximum intensities of the 420.2\,nm TE$_{00}$ and TE$_{01}$ modes at the surface of the rib are $1.2 \times 10^{10}$ W/m$^2$, while for the 980\,nm TE$_{00}$ mode it is $9 \times 10^{9}$ W/m$^2$. The trap minimum is located at $x_0=133.3$\,nm, where the maximum potential depth is $|U_{tot}(x_0,y)|/k_B=556.5$\,$\mu$K. The saddle points of the potential are located at  $x_s=98.0$\,nm and $y_s=\pm 1.05$\,$\mu$m, where the potential depth is $|U_{tot}(x_s,\pm y_s)|/k_B=243.2$\,$\mu$K. Therefore, the lateral depth of the potential is $\Delta |U_{tot}^{min}|/k_B=313.3$\,$\mu$K. The trap is characterized by $S=11.6$ and $\eta=0.74$, indicating good stability and efficiency. 

According to (\ref{EQ8}), the spontaneous scattering rate of photons of 420.2\,nm light at the minimum of the trap at $x=x_0$ is given by $\Gamma_{sc}(420.2\text{ nm})=298$\,s$^{-1}$. Note that this scattering rate cannot be estimated using a standard two-level approximation, because the excited $6P_{3/2}$ state can decay not only to $5S_{1/2}$ state, but also to other excited states of $^{87}$Rb \cite{23}. The scattering rate of the 980\,nm light is estimated to be  $\Gamma_{sc}(980\text{ nm})=0.78$\,s$^{-1}$. The heating rate of the trapped atoms is expected to be $dT/dt=124$\,$\mu$K/s. Therefore, the trapping time of atoms in this trap is about 2\,s.

In the absence of the TE$_{01}$ mode, the minimum trap depth is equal to 44.8\,$\mu$K. 

The spontaneous scattering rate can be decreased by increasing of the frequency detuning of the "blue" mode. For a wavelength of 420.2 \text{nm} and powers  P$_{\text{TE00}}(980 \text{nm})=6.3$\,mW, P$_{\text{TE00}}(419.3 \text{nm})=5.83$\,mW and P$_{\text{TE01}}(419.3 \text{nm})=7.34$\,mW, the minimum trap depth  $|\Delta U_{tot}^{min}|/k_B=22.8$\,$\mu$K, while the scattering rate is $\Gamma_{sc}(419.3\text{ nm})=1.9$\,s$^{-1}$ and the heating rate is $dT/dt=0.79$\,$\mu$K/s.

Therefore, the two-mode dual-resonant atomo-photonic waveguide is a working configuration and may offer advantages, as discussed in the next section.

\section{Discussion}

The two-mode, mono-resonant atomo-photonic waveguide operating at wavelengths of 698 nm and 1300 nm demonstrates excellent performance. Its potential depth is 99\,$\mu$K or above, depending on the relative portion of the TE$_{01}$ mode at 698\,nm. The reduced spatial anisotropy of the corresponding optical dipole potential is related to the large lateral size of the waveguide rib, which reduces the difference in lateral size between the 698\,nm and 1300\,nm TE$_{00}$ modes. This approach can certainly be applied not only to rubidium, but also to many other types of atoms.

The two-mode, dual-resonant atomo-photonic waveguide also demonstrates quite deep trapping potentials, albeit with larger spontaneous scattering rates.
One advantage of this configuration is that the two-mode, dual-resonant atomo-photonic waveguide can easily be transformed into a single-mode, mono-resonant waveguide. To achieve this, it is sufficient to switch the frequency of the blue-detuned two-mode mode to a single mode of a much lower frequency. This new mode is still blue-detuned, but now from the main atomic transition. For example, in the above case of the 420.3\,nm and 980\,nm modes, the 420.3\,nm double mode can be substituted with 640\,nm single mode without altering the geometry of the initial photonic waveguide. Such a switch to single-mode propagation may be important for the stability of the atomo-photonic waveguide parameters at their intersections, which are necessary for implementation of waveguide atom interferometers \cite{2}. Another advantage of using the high-frequency electron transitions of an atom is the much shorter wavelength of the corresponding light. This is beneficial for setting an optical lattice with a larger number of sites at the intersection of waveguides, enabling efficient coherent Bragg splitting of atoms between the two crossed waveguides \cite{28}.

The mode-selective coupling of the blue-detuned light to the TE$_{00}$ and TE$_{01}$ modes of the photonic waveguide can be achieved using directional couplers \cite{29} or properly tuning of other types of couplers.
To efficiently load such atomo-photonic waveguide, optical tweezers and moving optical lattices can be used \cite{30,31}. The planar geometry of all-optical atom chips suggests the possibility of implementing multiple atomtronic \cite{32} elements, such as coherent waveguide beam splitters \cite{28}, waveguide interferometers \cite{2}, crossed traps \cite{33} or their arrays.  

\section{Conclusion}

It has been demonstrated that two-mode suspended optical waveguides, including those in the dual-resonant regime, can significantly increase the potential depth of atomo-photonic waveguides. This could enable the coherent manipulation of ultra-cold atoms on all-optical atom chips, with potential applications in inertial sensing, atomic clocks and quantum computing.

\ack

The author is pleased to acknowledge the support of the National Physical Laboratory in the preparation of this manuscript.
Many thanks to Ian Hill, Sean Mulholland and Geoffrey Barwood for their comments and corrections to the paper.

\end{document}